\documentclass[acmsmall,screen,nonacm=true]{acmart}
\usepackage[utf8]{inputenc}
\usepackage{graphicx}
\usepackage{url}

\title{Remote Observation of Field Work on the Farm}
\author{Wendy Ju}
\email{wendyju@cornell.edu}
\affiliation{Cornell Tech}
\author{Ilan Mandel}
\email{im334}
\affiliation{Cornell Tech}
\author{Kevin Weatherwax}
\email{keweathe@ucsc.edu}
\affiliation{University of California, Santa Cruz}
\author{Leila Takayama}
\email{takayama@ucsc.edu}
\affiliation{University of California, Santa Cruz}

\author{Nikolas Martelaro}
\email{nikmart@cmu.edu}
\affiliation{Carnegie Mellon University}
\author{Denis Willett}
\email{deniswillett@cornell.edu}
\affiliation{Cornell AgriTech}

\begin{document}

\begin{abstract}
Travel restrictions and social distancing measures make it difficult to observe, monitor or manage physical fieldwork. We describe research in progress that applies technologies for real-time remote observation and conversation in on-road vehicles to observe field work on a farm. We collaborated on a pilot deployment of this project at Kreher Eggs in upstate New York. We instrumented a tractor with equipment to remotely observe and interview farm workers performing vehicle-related work. This work was initially undertaken to allow sustained observation of field work over longer periods of time from geographically distant locales; given our current situation, this work provides a case study in how to perform observational research when geographic and bodily distance have become the norm. We discuss our experiences and provide some preliminary insights for others looking to conduct remote observational research in the field.
\end{abstract}

\maketitle

\section{Introduction}

Farming is an increasingly difficult business, with an advancing median age of the domestic workforce, growing barriers to recruiting seasonal farm workers, and industrialization and consolidation creating greater competition in the market. 
This ``tough job'' has been made substantively tougher by the advent of COVID-19. 
Food producing farms and operations that care for animals have been exempted from the guidance for workforce reduction in most states, but travel restrictions and visa processing limitations have impacted the ability of many farms to hire H-2A workers who typically perform seasonal field work. \cite{USCIS2020}


Agricultural automation can help to change the labor equation in the fields, so that fewer people are able to farm more acres. \cite{geller2016farm} 
While agricultural robotics is advancing rapidly, surprisingly little research has been done to study the human-machine interaction between farm and agricultural workers and these new emerging systems. \cite{shamshiri2018research}
We have been studying human interaction with farm automation in the field---the literal field---by looking at how farm workers interact over the course of a growing season with existing farm automation. 
The novel methods we have devised to remotely study farm work can also help us to understand the changing nature of work on farms in the midst of the COVID-19 pandemic.


We have built and tested an interactive system for remote ethnography that allows us to perform longitudinal field studies of farm workers using automated farm vehicles. 
This system, FarmWoZ, is an adaptation of the Woz Way system which we originally developed to study interactions with automation in passenger vehicles. \cite{martelaro2017woz} 
Our system enables us to remotely observe how farm workers use current automation systems, supplement the audio and video data from our feed with sensor information, and record synchronized data in a way that allows ample post-analysis. 
To adapt to the challenges that the farming context brings to this style of research, we have adapted our system design so that it can be remotely operated for long-stretches of time. 
We have also built interfaces to better allow remote observers to contextualize what they are seeing and discussing with the farm workers. 
Our system is intended to help scientists, researchers, engineers and designers to better respond to the situated contexts of agricultural workers, over the course of the farming season, without having to be physically present on the farm. 

\section{Field Work to Date}
Kreher's Farm Fresh Eggs is a family-owned operation founded in 1924 in Upstate New York, located outside of Buffalo in Clarence, New York. 
This farm produces premium farm fresh eggs, organic and conventional produce crops such as corn, soybeans, and beets, as well as compost and fertilizer. 
As part of their overall operation, Kreher farms over 3,000 acres of certified organic crop land, both to feed their own chickens and for external sale. 
The Kreher family has close ties to Cornell; many of the family partners are alumni of the university. 

In the 2019-20 growing season, when we performed our test deployment of the FarmWoZ system, we interviewed farm managers about the use of automation on Kreher's Farms. 
As a mid-size farm, Kreher does not have the scale to transition their entire farm vehicle fleet to the latest automation systems, but they have been incorporating new automated equipment and machinery as old equipment is retired. 
With this gradual transition process, the farm scientists and workers experiment with different settings and applications to learn what works best for the specific crops and field conditions experienced on Kreher's crop land. 

The 2019 season turned out to be a challenging one; the long rain season in Spring 2019 prevented planting from occurring until late into the season \cite{rippey_usda_2019}. 
Many farmers did not plant crops at all, since the growing season would be so short that the return on investment was questionable \cite{the_counter_2020}. 
However, because the Kreher farm is in many ways a self-contained system, they needed to plant or otherwise they would not have feed for their chickens \cite{vogel_kwiatkowski_violanti_2011}. 

Our intent in the 2019 season was to witness a whole season; however, the incredible stress and pace of that particular season meant we were lucky to record 10 sessions of cultivating. 
Harvest turned out to be too hectic and stressful, and the workers we worked with understandably were resistant to having us back in the height of the harvest season to instrument the combines as we originally planned. 
However, we took a trip in early February this year to present our research in progress, and the crew at Kreher farms was excited to see all the nuances and details of their work that were captured.
They are looking forward to working together in the coming season, although it is currently unclear what shape that will take given current COVID-19 restrictions.

To our knowledge, this type of longitudinal examination of farm worker interaction is the first of its kind. One of the reasons this research is unique, and unusual, is that rural communities are distinct, and understudied in the United States \cite{hardy2019designing}. 
Another reason, however, is that field work is expensive and difficult if researchers are not located near the subject of the research; our system helps to bridge this gap. 
In a moment where travel is restricted, and social distancing is the norm, systems that enable sustained remote research are vital to producing an understanding of contexts that demand design and engineering.







\begin{figure}[t]
  \centering
  \includegraphics[width=\textwidth]{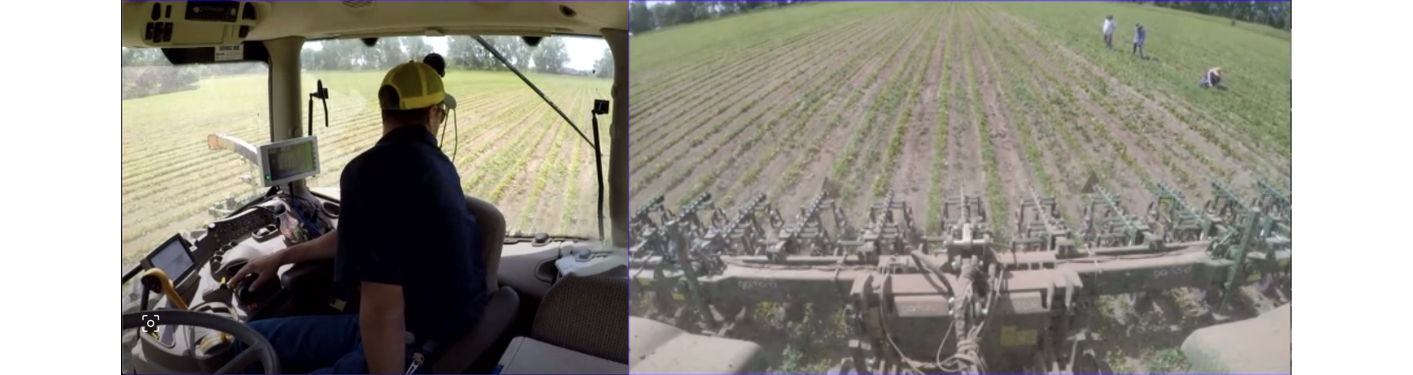}
  \caption{(Left) An operator in a semi-automated tractor. (Right) A RoboCrop Cultivator Head running in a field of beets. Note the H2A workers outside of the path of the tractor on the right.}
\end{figure}

\section{Prior Work}
While notable work in remote ethnography has occurred in the social sciences \cite{postill2016doing}, much of the prior work is premised on digital technologies such as video conferencing or mobile phones, and hence has developed amidst the field of human-computer interactions \cite{kushniruk2015video,pink2016digital}. 
Early systems focused on extending the experience sampling method \cite{csikszentmihalyi_ecology_1977} as a technique to evaluate and improve ubiquitous computing applications \cite{consolvo_using_2003}. 
Froehlich's MyExperience, for example, enables designers to survey users after specific interactions with their device \cite{froehlich_myexperience:_2007}. 
One example had users rate the call quality after a mobile phone call on the phone itself. 
Carter et al.'s Momento \cite{carter_momento:_2007} was developed to help designers better understand the in-world context in which people were using mobile computing applications. 
Designers receive notifications around specific trigger events notifying of a participant's mobile application use, in real-time. They can subsequently interact with the user over multimedia messaging by sending questions and requesting photos or videos of the user's environment. 


In discussing the future of remote interaction design tools, Crabtree et al.\ point out that many aspects of interaction in ubiquitous environments are invisible and fragmented, and that \textit{``there is a strong need to enhance observation in these environments, making the invisible visible and reconciling the fragments to permit coherent description.''} \cite{crabtree_supporting_2006} 
They champion the combined use of video and system data in remote ethnography to enable sensemaking of increasingly computational user experiences. 
Hence, the WozWay system \cite{martelaro2017woz} that FarmWoZ is derived from provides high quality video, multiple camera angles to establish context in and out of the cabin, and also provides map data to remote interactants to help them localize what they are seeing when they interact with drivers.

\section{Methods}

The objectives for our novel research were to:

1) Develop a system that enables us to perform remote field studies of people interacting with farm automation. 
We aimed to design a system that is inexpensive, remotely operable, remotely deployable, and easy to use, so as to maximize adoption by other researchers.


2) Perform field studies of people interacting with automated farm equipment throughout the growing and harvest seasons. 
This was an initial foray into establishing key issues in human-machine interaction for farm automation systems based on empirical longitudinal study.

Here we describe each of these approaches and methods in greater detail:
\subsection{FarmWoZ: An Interactive System for Remote Ethnography}

\begin{figure}[t]
  \centering
  \includegraphics[width=\textwidth]{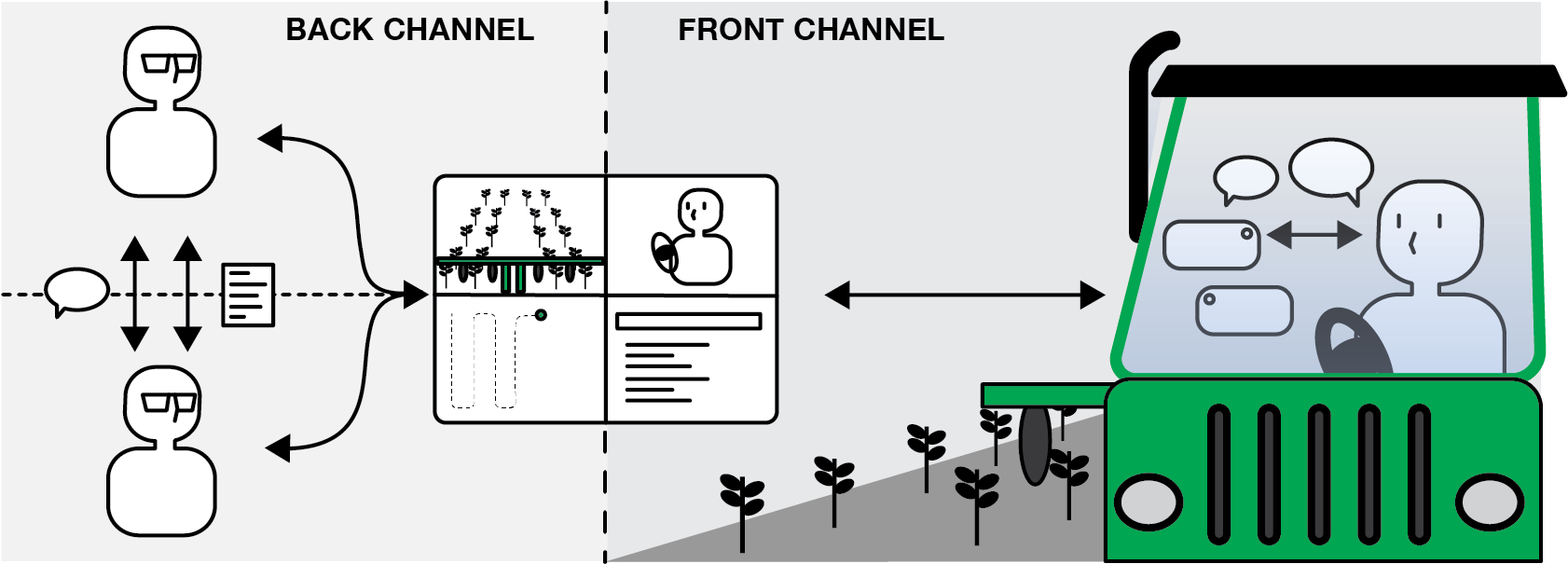}
  \caption{System diagram showing remote and in-field parts of our remote ethnography system.}
  \label{fig:system_diagram}
\end{figure}

\subsubsection{Function} 
The FarmWoZ system, derived from the WoZ Way system \cite{martelaro2017woz}, allows researchers to watch the real-time tractor cabin experience via high fidelity video and audio, and simultaneously receive meta-data about the drive such as real-time map or sensor information. 
The researcher can also interact with the operator, asking questions by using a text-to-speech messaging system.
A system diagram for our FarmWoZ Interactive Remote Ethnography system is shown in \ref{fig:system_diagram}.


At present, our current system costs around \$5000 and fits in a carry-on sized Pelican case. 
It includes three GoPro cameras, a video mixer, an HDMI video input device, a laptop computer, a cellular router, a GPS/IMU sensor unit for location tracking, and a GoalZero Yeti 150 portable battery power station. 



\subsubsection{Data Streaming and Capture}
FarmWoZ has been designed to provide the researcher more contextual information than potentially possible if the researcher were merely riding along, such as a live stream of map and sensor data and multiple camera views.
In our deployment we used three cameras, one to view the tractor operator, one to view out the front of the tractor, and one to view the cultivator equipment pulled behind the tractor.
We also use a GPS to follow the vehicle's location through the field. 

Video and audio are streamed from the cabin using a video chat client. 
Location data and the text-to-speech messages are streamed through a separate, centralized data server. 
The Wizard interface shows a live map, 


\subsubsection{Deployment Vehicle}
We deployed the FarmWoZ system in a John Deere tractor with a robotic cultivator tool attached.
The tractor was equipped with self-steering capabilities powered by a Trimble precision GPS system.
The self-steering system helps maintain straight driving at a set speed but does not automatically turn the tractor around at the end of a field row.
The Robocrop cultivator tool,  was equipped with a dual-camera computer vision system that helped to maintain it's position equally between the plant rows.
The cultivator head makes fine adjustments left and right to maintain perfectly straight rows but does not lift itself if there are obstacles such as rocks in the rows.

\subsubsection{Analysis}
\begin{figure}[t]
  \centering
  \includegraphics[width=\textwidth]{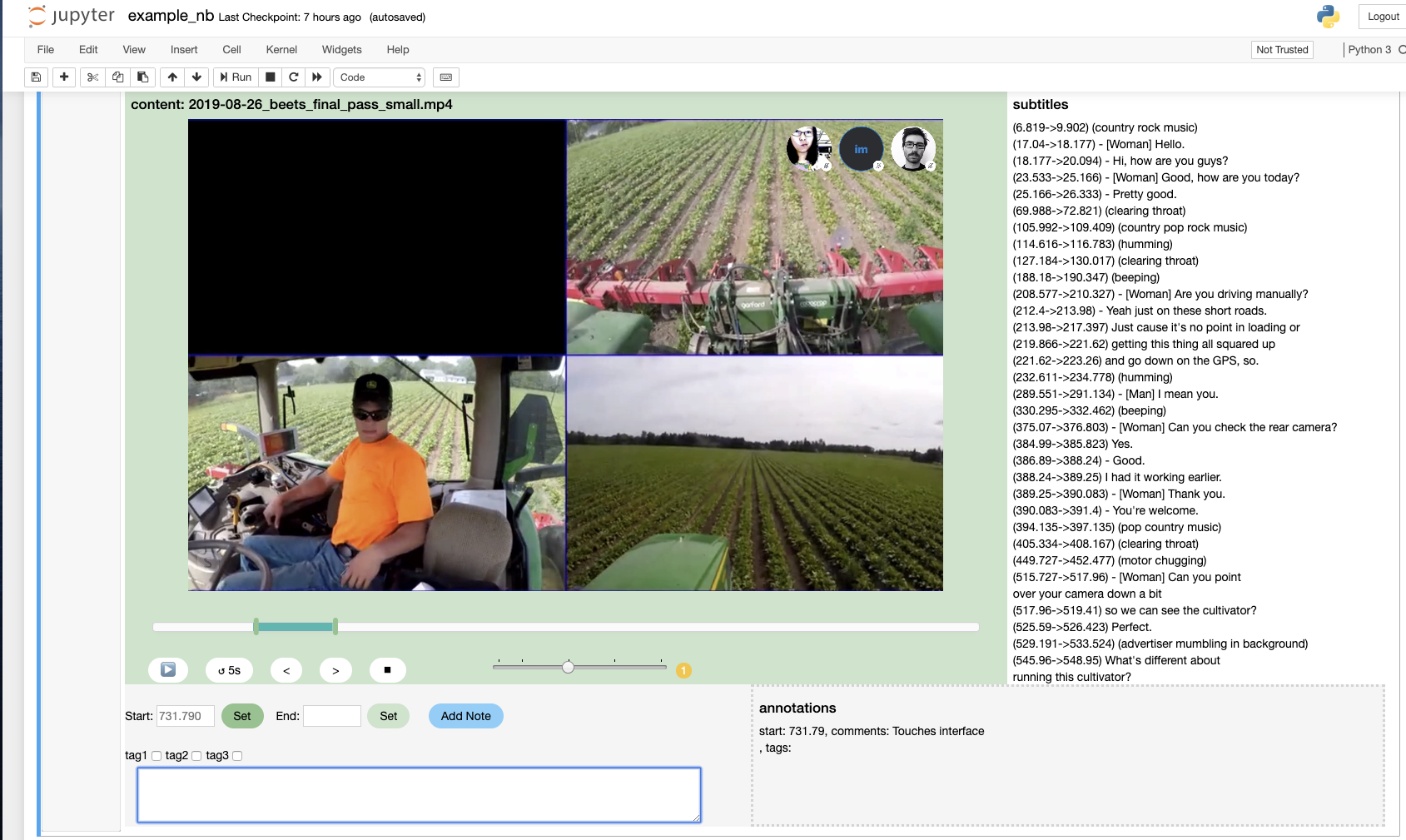}
  \caption{Prototype interface for multi-researcher collaborative analysis of captured video.}
  \label{fig:analysis}
\end{figure}

We are developing tools to allow multiple researchers to perform post-facto analysis of the data collected by our FarmWoZ system. 
We believe this to be important for the interdisciplinary collaborative aspects of our work, as no one person can see all the important things that need to be recognized in the footage collected. 
The system is built as a Jupyter notebook to make it easier to use the system as the first step in building machine learning tools to help automatically recognize and parse features corresponding to researcher-generated labels.
A prototype for this analysis interface is shown in Figure \ref{fig:analysis}.

\section{Preliminary findings and discussion} 

The outcomes of our work to date are:

1) An interactive system and method to support remote longitudinal study of farm work.

2) A deeper understanding of the constraints and opportunities to improve human interaction with automated farm vehicles.

For the purposes of addressing the larger question of how real-time remote interaction prototyping and observation technologies can help enable remote observation, monitoring or management in the time of COVID-19, we will focus our findings and discussions on the first outcome. 




\begin{figure}[t]
\centering
  \includegraphics[width=1\columnwidth, trim=0 500 0 0, clip]{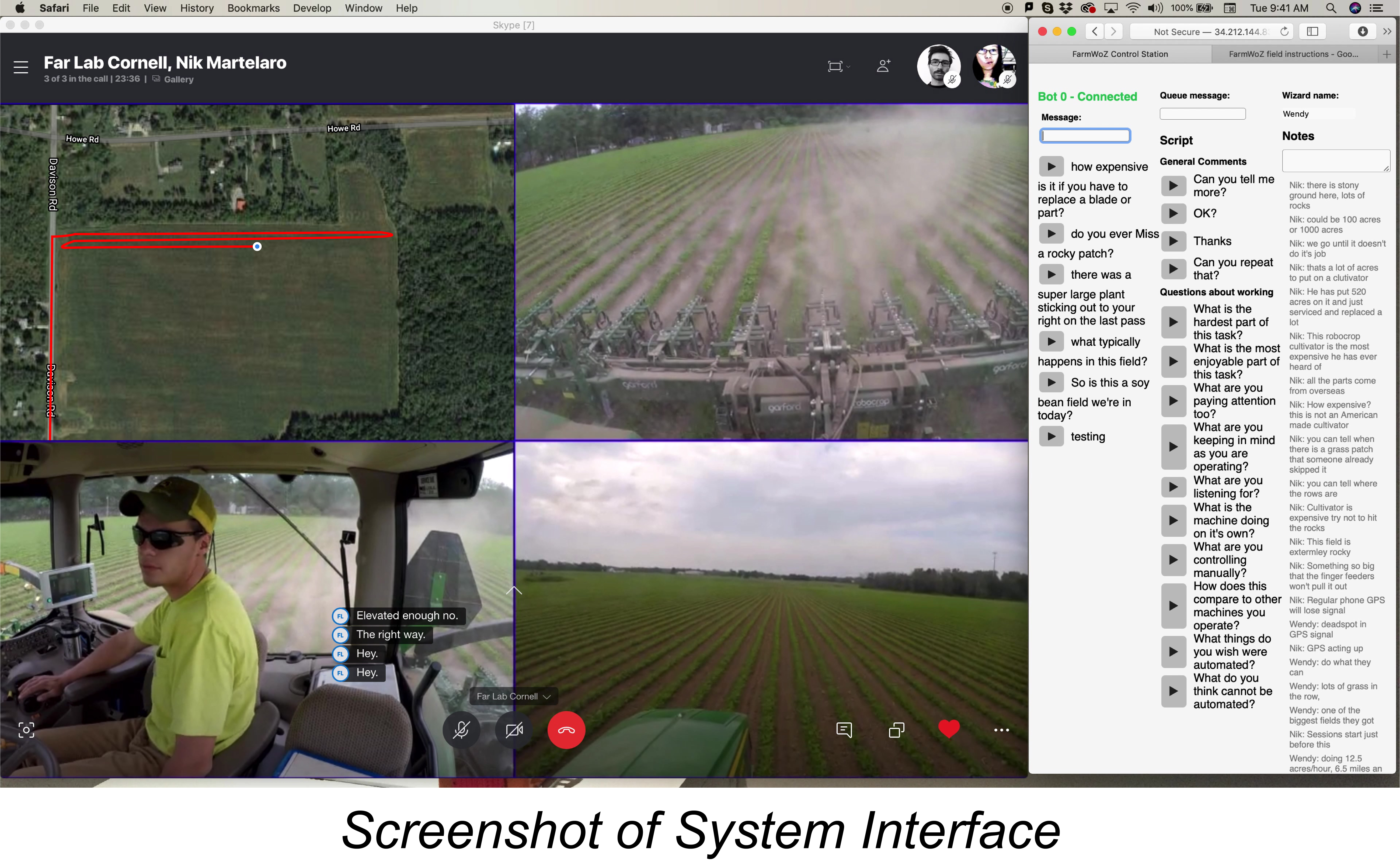}
  \caption{Screenshot of an pilot data and control interface. }~\label{fig:wizInterface}
\end{figure}

\subsection{Preparation and setup as critical path} 
One critical aspect of deployments such as ours is the set up of the equipment used to capture, record and transmit the video and location data used in our study, as well as the substantial amount of fixturing and cabling needed to connect and power the equipment. 
In this deployment, we had a day on site in late February 2019, where we talked to the people we would be working with. 
Discussions of which vehicle to instrument, who we would work with on a day to day basis, what the key factors were to observe were had in that meeting, and the researchers climbed into five different farm vehicles to assess the viability of each. This site visit also enabled grounding conversations about how often each vehicle was used, by which workers, and during which part of the growing season. 
This aspect of the deployment is difficult to replicate without a visit to the physical site. 

The actual installation of the equipment took the better part of a day in May 2019, with personnel that were both on- and off- site, in part because of the needs 1) to work out the physical location of the different part components, 2) to check that all the parts received enough power when the vehicles were turned on and off, 3) to figure out whether we would need to externally charge the portable battery we had for the system, and 4) to verify that we had sufficient cellular coverage to make the remote connection viable. 
We also spent a lot of time on site labelling all of the parts and creating a detailed guidebook for people on-site to use should they need to service or move any of the equipment.

Without the initial scouting visit, a person who had prior knowledge of the context of deployment and the technologies that would be used would be needed to understand where and how to set up the instrumentation. This is a significant limitation and could hamper or prevent system deployment in a COVID-19 era research setting.

\begin{figure}[t]
  \centering
  \includegraphics[width=.8\textwidth]{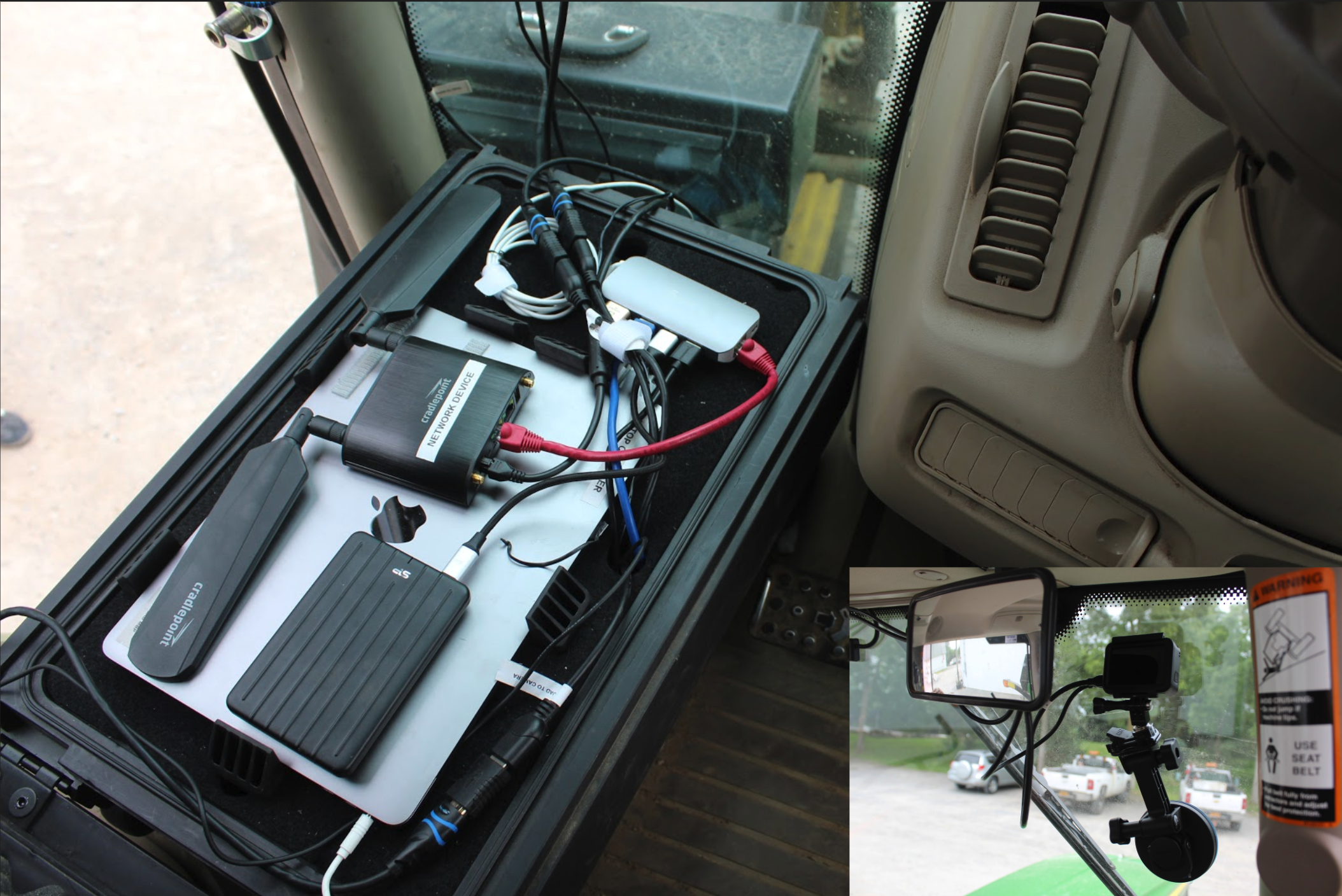}
  \caption{Original equipment installed in farm vehicle.}
  \label{fig:pelican_case}
\end{figure}

\begin{figure}[t]
  \centering
  \includegraphics[width=.8\textwidth]{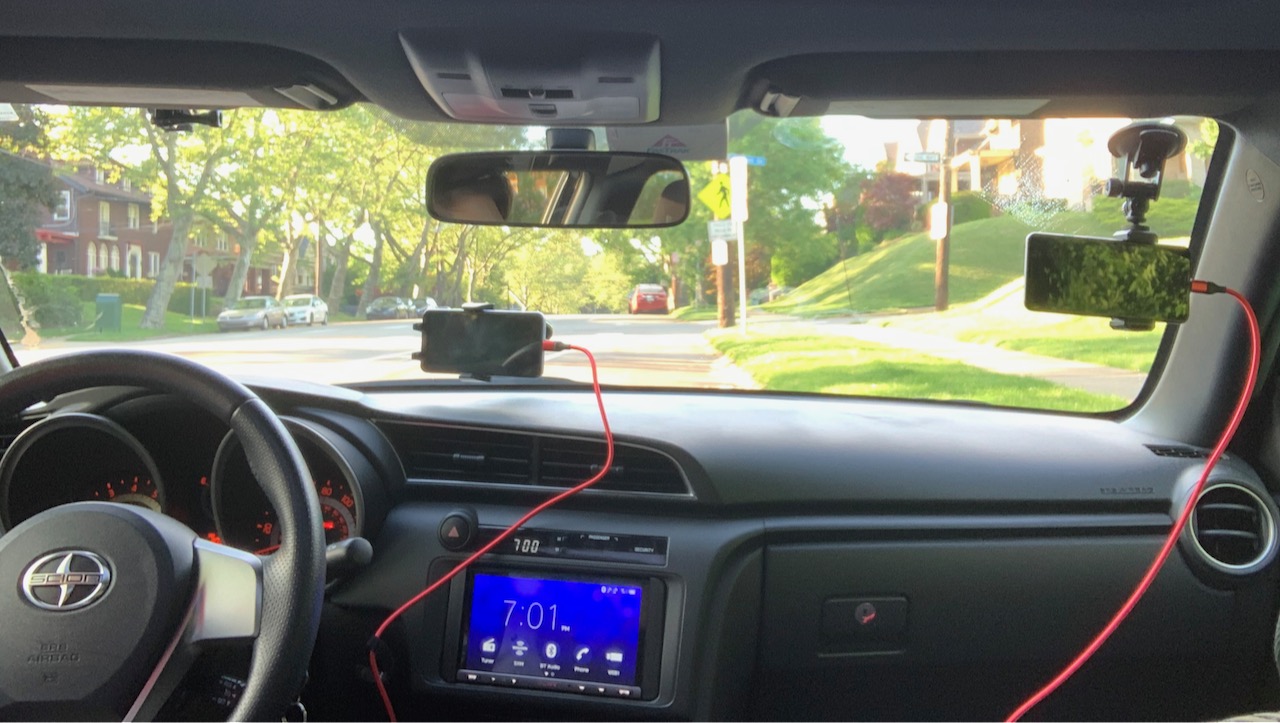}
  \caption{Test deployment of next generation FarmWoZ instrumentation, in a car, using two phones for driver and front road view.}
  \label{fig:two-phone-setup}
\end{figure}

For this reason, both to ease the need for complex system knowledge and to make the equipment installation and maintenance lower, we have substantially reduced the infrastructure needed to support the remote observation. 
The original equipment for the deployment is shown in Figure \ref{fig:pelican_case}. 
While this setup allows us a high data-rate high video quality recording, with the potential for back-up video recording on the computer hard disk, we realized that a simpler setup would be worth considerable trade-offs. 
For this coming season's deployment of this research, we have developed a new version of the FarmWoZ system to run on cellular phones running the Android operating system, which can be mounted in the farm vehicle using suction mounts, and powered using standard 12v in-vehicle charging apparatus for phones. (See Figure \ref{fig:two-phone-setup}.) 

With this set up, we are able to have multiple views of the in-vehicle work setting, to perform text-to-speech interaction, and to gather GPS and phone-based sensor information which can be relevant to establishing a larger sense of context.
We have developed automated mechanisms to allow remote researchers to remotely set up all aspects of the system after somebody has mounted the phones and powered them on, so long as the phones are in cellular range. 
This set up is less expensive from an equipment perspective (~\$750 USD vs. \$5000 USD) but costs more to run in an on-going basis, because of the costs of the cellular data plans for each phone (roughly \$50 USD per phone per month of deployment). 
A custom system could be built using an embedded computer such as the NVIDIA Jetson Nano, but such a system would not be as familiar and easy to use for the local participants as the cellular phones.
Overall, we would recommend that researchers looking to deploy remote observation systems consider how much complexity their participants should manage and work to build systems that are easy to set up and remotely manage.

\subsection{System Operations}
Each of our FarmWoZ observation sessions took some coordination with the person who was using the tractor, because they had to turn the system on. 
The overall boot-up process did not take long---we estimate 5 minutes turning on the computer, camera, and wireless network---but it did cause a constraint because the person who did the boot-up was ideally the person who was initially trained.
The boot-up time was also a challenge because it made workers not want to set things up when work on the farm was a little more hectic, complicated or stressful than normal, which occurred numerous times during the growing season we observed.

One question about the system we designed and tested in our pilot deployment was whether we needed all the cameras we deployed. 
During the observations, we did find that all of the camera views we set up in the farm vehicle were necessary. 
Our system had the capability to connect up to four cameras; we set up three, one of the farm worker, one out the front of the farm vehicle, and one out the back. 
We viewed this as likely a super set of what was needed, but we thought we would see if any of the views would prove to be unnecessary. 
As it was, all three camera views were important. 
The view of the farm vehicle operator was important, because it would have been difficult to ascertain what the worker was doing without that view. 
It would be possible to ask, but understanding when we could ask about things without distracting the worker would have been difficult. 
The view out the front of the tractor was critical to understand the context of where the vehicle was driving, and to see what the worker was attending to. 
This view also helped us to see when the tractor was at the end of a row, or when there was an obstacle in the path, or what the weather looked like. 
The view out of the rear of the tractor was important to understanding the actual work that was being performed; we were able to see the rows being cultivated, and how the robotic cultivator was or wasn't performing its task.

One naive question that anyone might have of the situation we were observing is what it is that the farm worker has to do when they are in a vehicle that is largely self-driven; there is GPS navigation to pinpoint the location of the tractor, and the automated Robocrop cultivator uses computer vision to line up the cultivator head with the rows of the crop. 
However, what we witnessed by watching the cultivation activity showed that the farm worker needs to constantly monitor all of the automated systems and perform small corrections or overrides to keep the system working smoothly. 
We noted that at times, the farm worker looked like he was getting an upper body workout from needing to turn forward and back to see how things were going. 
For instance, one time, our worker said "ROCK" and put his hand on the cultivator controls. 
Looking out into the field, we did not see a rock, but we did see a patch of the field which was uncultivated, missing the visible rows that were in the other parts of the field. 
When we asked about it, he said that the uncultivated patch indicated where other people had also lifted the cultivator head, and usually marked areas where someone had previously seen a rock. He explained that the automated equipment was so expensive that repairs and downtime were extremely costly; it was better to assume there was a rock where an uncultivated patch emerged than to risk damage. 
The amount and type of activity the farm worker did changed over the course of the season. 
Early on, there was a lot he needed to do to make sure that the automated cultivator lined itself up properly, and to profile portions of the field where the GPS signals might be weak and the vehicle would lose its location. 
Later in the season, there was more growth for the automated cultivator to see, and the driver had built up knowledge about where the system could be trusted to drive itself.
During our observations sessions, we could see and hear that the driver had more time to check his cell phone or relax in between moments when he was looking to see if the crops were growing well or if weeds were growing faster or slower than expected.

Based on this experience, we felt that all the cameras we originally set up were necessary to monitor activities, to contextualize discussions, and to understand what worker was doing and why throughout the growing season. 
Substantially more was learned by being able to talk to the farm worker by watching activity and asking about it in real-time than would have been learned in post-facto interviews, even if we had performed multiple interviews throughout the season. 
Our experiences from the FarmWoZ operations strongly suggests the value of performing remote longitudinal studies even under work-from-home restrictions, and also validates the importance of having multiple cameras to make that remote observation and real-time interaction as rich as can be for the remote parties.


\subsection{Missed moments}
We would be remiss if we did not also mention the numerous ``missed moments'' that occurred in our pilot deployment. 
Very often, we would learn of things that we would have liked to observe and ask questions about during our conversations with the worker. 

For example, one time the worker mentioned that it was good that we stopped when we did in the previous session because shortly thereafter the cultivator broke a tine, and the tractor had to be stopped and other workers had to come to repair the tractor. 
He seemed surprised when we reacted in disappointment to having missed that event; to his mind, the breakage and repair of farm equipment represented an undesirable stoppage in farm work, and not part of the work that we would be interested in witnessing and discussing. 
He laughingly promised that we would hopefully see another breakdown that he hoped would never happen. 
But, we never did get to see such an event. 
After those conversations, however, evidence of repair as a core activity in the farm work emerged in lots of places; for example, the worker had a bag of repair equipment that usually sat on the jumpseat where our recording equipment was installed, and we discussed whether that equipment was safe to sit on top of the Pelican case for our equipment.

Another activity we missed seeing was night work. Because of the aforementioned short season, the farm workers often worked into the darkness of the night to plant and cultivate fields. The worker we were talking to mentioned that this was a little bit dangerous, because the lower visibility made it harder to see if the tractor and the cultivator were doing exactly what they should, and the chances of damaging equipment by running into a rock or wild animal in the field increased. Once we expressed interest in seeing night work, numerous attempts were made to enable such a session. However, usually the need to perform night work also intersected with stressful moments in the farm work, and for a number of contributing factors, we never had a session of night time activity.

The stress of the difficult season also influenced the timing of our work, which also influenced what we could witness. Different crops are planted at different times on the farm; since our FarmWoZ equipment was set up in a tractor with a cultivator, we primarily witnessed cultivation of already-planted crops. It seemed possible that we might have been able to see the planting of other, later-installed plants, but the effort to move the instrumentation seemed to be too great to overcome. We mostly watched the cultivation of corn and beets, and it was difficult to know what we might have missed if we were able to see farm activity on a wider range of crops, or interview a wider range of workers.

We believe there are two parts to the aspect of missing moments in the remote observation of field work. One part has to do with scale; there is only so much we can witness through the instrumentation of one vehicle, and interacting with one worker. A more comprehensive view of our subject matter is possible if we can replicate our system and install it in more vehicles; the lower cost and complexity of our new system should help to make that possible.

The other aspect to the missing moments has to do with the participant engagement. It takes investment of time and effort for our participants to interact with us during the course of their work through FarmWoZ, even if we are trying to make the most of relatively relaxed moments in in-vehicle farm activity. Many of the ``missed moments'' we most hoped to see were also moments of high stress and uncertainty. Reducing the complexity of our system certainly helps with that, but, as we will discuss later, it is also important to provide the participants more transparency into the value of the work being observed to help motivate the effort required.

\subsection{Direct interaction vs. long-term monitoring}

It is clear that many of the previously mentioned ``missed moments'' could have been captured if we had a passive ``always on'' system that was constantly recording activity that was happening in and around the farm vehicle. To our mind, this brings up a crucial distinction between longitudinal study through the direct interaction made possible by FarmWoZ and long-term activity monitoring. In our FarmWoZ work, the researchers are remotely witnessing and experiencing the work activity in real-time, and are able to ask about it; the worker is able to describe the context and the in-the-moment thinking around the activity. The back and forth keeps the worker aware that his activities are being observed, and he is able to adapt his behavior to that context, to point out factors that cannot be seen on camera, or to voice in-the-moment uncertainties before the influence of what happens later colors recollections of those thoughts. Because we have not performed long-term passive monitoring as a research method, we are not able to make direct comparisons of the approaches on outcomes and findings. That said, we feel that the FarmWoZ system does a better job of capturing the ethnographic and design research moments that would also be captured by longitudinal ethnographic study than passive recording and retrospective interviewing would.

\subsection{Ethics, Transparency and Feedback}
Because we are observing and recording people in their place of work, there are inherently ethical issues around surveillance and job risk. \cite{zuboff2019age} We feel that the fact that we are not always recording, and the fact that we are actively interacting with the participants when we are recording, helps the participant to be aware of our presence and actions, and allows them to modify their behavior or negotiate our own activity. We are also selective about the moments that we choose to show, to make sure they always reflect our intent which is to make better systems for people working on farms to use.

Longitudinal studies involve considerable investment and work on the part of the participants.
Our primary tractor operator acted as a research participant and would help us to debug the FarmWoZ system when the communications broke down.
This included re-routing power cables, restarting software, and debugging wireless connections.
While our participant was amenable to helping, it did cut into their working hours and at times they did not have the time to work with us.
Many of these challenges can be solved through more robust systems, however we also found that having more communication with the farm team about our research helped us to build more rapport and investment, leading to more engagement and a willingness to work with us.

After our pilot deployment, we traveled to the farm and presented reflections on how the deployment went, key moments from our observations, and our plans for moving the project forward.
This meeting lead to a robust discussion with the entire farm team and increased interest in the project.
After discussing our results with the team there was a renewed interest to set up the FarmWoZ system in different equipment and continue observations during different parts of the farming season. 
This transparency also may have an effect on the research, as the participants may alter their behavior based on the awareness that they are being studied, and may also point the observations towards things they want the researchers to attend to. At the same time, this transparency and feedback can help make workers more aware of the intent of the overall research efforts, and increase their willingness to make concessions and do extra work to accommodate research activity. 
While it is obvious that forming good relationships is important for any longitudinal ethnography, being remote makes this more challenging.
When researching in the field in person, research participants can more directly see the products of the research work.
Researchers can also form more personal connections with participants, helping to increase trust.
Being remote makes this rapport building more difficult due to reduced interaction time.
Additionally, while behind the FarmWoZ system, we as researchers are represented by a computer voice, rather than as individuals.
Having frequent phone calls with our primary operator helped to maintain the connection with our team as people.

As we move forward, we plan to have more engagements where we show what we are learning to the team and coordinate on how to improve our system and interactions in-cabin.
Other researchers looking to use remote ethnography methods should consider how they can build and maintain rapport with their participants over time.
Minimizing the complexities of remote interaction can help minimize wearing participants out by reducing their required effort.
However, coordinating remote research always has added complexities.
Making participants aware of how they are positively contributing to your work and involving them in the process can help to maintain engagement over long periods of time.

\section{Conclusion} 
We hope that the novel research that we have performed in conducting remote observation of fieldwork on the farm can inform discussions of the new future of work in the COVID-19 area. The work we performed on the FarmWoZ project speaks to possible ways that designers, researchers and managers can remotely observe and interact with people in the field. We strongly believe this connection is critical to making human-centered systems even under pandemic circumstances.

\bibliographystyle{ACM-Reference-Format}
\bibliography{farmwoz}


\begin{thebibliography}{17}


\ifx \showCODEN    \undefined \def \showCODEN     #1{\unskip}     \fi
\ifx \showDOI      \undefined \def \showDOI       #1{#1}\fi
\ifx \showISBNx    \undefined \def \showISBNx     #1{\unskip}     \fi
\ifx \showISBNxiii \undefined \def \showISBNxiii  #1{\unskip}     \fi
\ifx \showISSN     \undefined \def \showISSN      #1{\unskip}     \fi
\ifx \showLCCN     \undefined \def \showLCCN      #1{\unskip}     \fi
\ifx \shownote     \undefined \def \shownote      #1{#1}          \fi
\ifx \showarticletitle \undefined \def \showarticletitle #1{#1}   \fi
\ifx \showURL      \undefined \def \showURL       {\relax}        \fi
\providecommand\bibfield[2]{#2}
\providecommand\bibinfo[2]{#2}
\providecommand\natexlab[1]{#1}
\providecommand\showeprint[2][]{arXiv:#2}

\bibitem[\protect\citeauthoryear{Bloch}{Bloch}{2019}]%
        {the_counter_2020}
\bibfield{author}{\bibinfo{person}{Sam Bloch}.}
  \bibinfo{year}{2019}\natexlab{}.
\newblock \bibinfo{title}{\#noplant19: Unrelenting rain forces Midwest farmers
  to make a painful choice-plant or wait until next year}.
\newblock
\newblock
\urldef\tempurl%
\url{https://thecounter.org/climate-change-corn-planting-rain-midwest/}
\showURL{%
\tempurl}


\bibitem[\protect\citeauthoryear{Carter, Mankoff, and Heer}{Carter
  et~al\mbox{.}}{2007}]%
        {carter_momento:_2007}
\bibfield{author}{\bibinfo{person}{Scott Carter}, \bibinfo{person}{Jennifer
  Mankoff}, {and} \bibinfo{person}{Jeffrey Heer}.}
  \bibinfo{year}{2007}\natexlab{}.
\newblock \showarticletitle{Momento: {Support} for {Situated} {Ubicomp}
  {Experimentation}}. In \bibinfo{booktitle}{\emph{Proceedings of the {SIGCHI}
  {Conference} on {Human} {Factors} in {Computing} {Systems}}}
  \emph{(\bibinfo{series}{{CHI} '07})}. \bibinfo{publisher}{ACM},
  \bibinfo{address}{New York, NY, USA}, \bibinfo{pages}{125--134}.
\newblock
\showISBNx{978-1-59593-593-9}
\urldef\tempurl%
\url{https://doi.org/10.1145/1240624.1240644}
\showDOI{\tempurl}


\bibitem[\protect\citeauthoryear{Consolvo and Walker}{Consolvo and
  Walker}{2003}]%
        {consolvo_using_2003}
\bibfield{author}{\bibinfo{person}{Sunny Consolvo} {and}
  \bibinfo{person}{Miriam Walker}.} \bibinfo{year}{2003}\natexlab{}.
\newblock \showarticletitle{Using the {Experience} {Sampling} {Method} to
  {Evaluate} {Ubicomp} {Applications}}.
\newblock \bibinfo{journal}{\emph{IEEE Pervasive Computing}}
  \bibinfo{volume}{2}, \bibinfo{number}{2} (\bibinfo{year}{2003}),
  \bibinfo{pages}{24--31}.
\newblock
\showISSN{1536-1268}


\bibitem[\protect\citeauthoryear{Crabtree, Benford, Greenhalgh, Tennent,
  Chalmers, and Brown}{Crabtree et~al\mbox{.}}{2006}]%
        {crabtree_supporting_2006}
\bibfield{author}{\bibinfo{person}{Andy Crabtree}, \bibinfo{person}{Steve
  Benford}, \bibinfo{person}{Chris Greenhalgh}, \bibinfo{person}{Paul Tennent},
  \bibinfo{person}{Matthew Chalmers}, {and} \bibinfo{person}{Barry Brown}.}
  \bibinfo{year}{2006}\natexlab{}.
\newblock \showarticletitle{Supporting {Ethnographic} {Studies} of {Ubiquitous}
  {Computing} in the {Wild}}. In \bibinfo{booktitle}{\emph{Proceedings of the
  6th {Conference} on {Designing} {Interactive} {Systems}}}
  \emph{(\bibinfo{series}{{DIS} '06})}. \bibinfo{publisher}{ACM},
  \bibinfo{address}{New York, NY, USA}, \bibinfo{pages}{60--69}.
\newblock
\showISBNx{978-1-59593-367-6}
\urldef\tempurl%
\url{https://doi.org/10.1145/1142405.1142417}
\showDOI{\tempurl}


\bibitem[\protect\citeauthoryear{Csikszentmihalyi, Larson, and
  Prescott}{Csikszentmihalyi et~al\mbox{.}}{1977}]%
        {csikszentmihalyi_ecology_1977}
\bibfield{author}{\bibinfo{person}{Mihaly Csikszentmihalyi},
  \bibinfo{person}{Reed Larson}, {and} \bibinfo{person}{Suzanne Prescott}.}
  \bibinfo{year}{1977}\natexlab{}.
\newblock \showarticletitle{The ecology of adolescent activity and experience}.
\newblock \bibinfo{journal}{\emph{Journal of Youth and Adolescence}}
  \bibinfo{volume}{6}, \bibinfo{number}{3} (\bibinfo{date}{Sept.}
  \bibinfo{year}{1977}), \bibinfo{pages}{281--294}.
\newblock
\showISSN{0047-2891, 1573-6601}
\urldef\tempurl%
\url{https://doi.org/10.1007/BF02138940}
\showDOI{\tempurl}


\bibitem[\protect\citeauthoryear{Froehlich, Chen, Consolvo, Harrison, and
  Landay}{Froehlich et~al\mbox{.}}{2007}]%
        {froehlich_myexperience:_2007}
\bibfield{author}{\bibinfo{person}{Jon Froehlich}, \bibinfo{person}{Mike~Y.
  Chen}, \bibinfo{person}{Sunny Consolvo}, \bibinfo{person}{Beverly Harrison},
  {and} \bibinfo{person}{James~A. Landay}.} \bibinfo{year}{2007}\natexlab{}.
\newblock \showarticletitle{{MyExperience}: {A} {System} for in {Situ}
  {Tracing} and {Capturing} of {User} {Feedback} on {Mobile} {Phones}}. In
  \bibinfo{booktitle}{\emph{Proceedings of the 5th {International} {Conference}
  on {Mobile} {Systems}, {Applications} and {Services}}}
  \emph{(\bibinfo{series}{{MobiSys} '07})}. \bibinfo{publisher}{ACM},
  \bibinfo{address}{New York, NY, USA}, \bibinfo{pages}{57--70}.
\newblock
\showISBNx{978-1-59593-614-1}
\urldef\tempurl%
\url{https://doi.org/10.1145/1247660.1247670}
\showDOI{\tempurl}


\bibitem[\protect\citeauthoryear{Geller}{Geller}{2016}]%
        {geller2016farm}
\bibfield{author}{\bibinfo{person}{Tom Geller}.}
  \bibinfo{year}{2016}\natexlab{}.
\newblock \showarticletitle{Farm automation gets smarter}.
\newblock \bibinfo{journal}{\emph{Commun. ACM}} \bibinfo{volume}{49},
  \bibinfo{number}{11} (\bibinfo{year}{2016}), \bibinfo{pages}{18--19}.
\newblock


\bibitem[\protect\citeauthoryear{Hardy, Phelan, Vigil-Hayes, Su, Wyche, and
  Sengers}{Hardy et~al\mbox{.}}{2019}]%
        {hardy2019designing}
\bibfield{author}{\bibinfo{person}{Jean Hardy}, \bibinfo{person}{Chanda
  Phelan}, \bibinfo{person}{Morgan Vigil-Hayes}, \bibinfo{person}{Norman~Makoto
  Su}, \bibinfo{person}{Susan Wyche}, {and} \bibinfo{person}{Phoebe Sengers}.}
  \bibinfo{year}{2019}\natexlab{}.
\newblock \showarticletitle{Designing from the rural}.
\newblock \bibinfo{journal}{\emph{interactions}} \bibinfo{volume}{26},
  \bibinfo{number}{4} (\bibinfo{year}{2019}), \bibinfo{pages}{37--41}.
\newblock


\bibitem[\protect\citeauthoryear{Kushniruk and Borycki}{Kushniruk and
  Borycki}{2015}]%
        {kushniruk2015video}
\bibfield{author}{\bibinfo{person}{Andre~W Kushniruk} {and}
  \bibinfo{person}{Elizabeth~M Borycki}.} \bibinfo{year}{2015}\natexlab{}.
\newblock \showarticletitle{Video Analysis and Remote Digital Ethnography:
  Approaches to understanding user perspectives and processes involving
  healthcare information technology.}
\newblock In \bibinfo{booktitle}{\emph{Studies in health technology and
  informatics}}. \bibinfo{publisher}{IOS Press}, \bibinfo{address}{Amsterdam,
  NL}, \bibinfo{pages}{206--216}.
\newblock


\bibitem[\protect\citeauthoryear{Martelaro and Ju}{Martelaro and Ju}{2017}]%
        {martelaro2017woz}
\bibfield{author}{\bibinfo{person}{Nikolas Martelaro} {and}
  \bibinfo{person}{Wendy Ju}.} \bibinfo{year}{2017}\natexlab{}.
\newblock \showarticletitle{WoZ Way: Enabling Real-Time Remote Interaction
  Prototyping \& Observation in On-Road Vehicles}. In
  \bibinfo{booktitle}{\emph{Proceedings of the 2017 ACM Conference on Computer
  Supported Cooperative Work and Social Computing}} (Portland, Oregon, USA)
  \emph{(\bibinfo{series}{CSCW '17})}. \bibinfo{publisher}{Association for
  Computing Machinery}, \bibinfo{address}{New York, NY, USA},
  \bibinfo{pages}{169–--182}.
\newblock
\showISBNx{9781450343350}
\urldef\tempurl%
\url{https://doi.org/10.1145/2998181.2998293}
\showDOI{\tempurl}


\bibitem[\protect\citeauthoryear{Pink}{Pink}{2016}]%
        {pink2016digital}
\bibfield{author}{\bibinfo{person}{Sarah Pink}.}
  \bibinfo{year}{2016}\natexlab{}.
\newblock \showarticletitle{Digital ethnography}.
\newblock In \bibinfo{booktitle}{\emph{Innovative methods in media and
  communication research}}. \bibinfo{publisher}{Springer International
  Publishing}, \bibinfo{address}{Switzerland}, \bibinfo{pages}{161--165}.
\newblock
\showISBNx{9783319407005}


\bibitem[\protect\citeauthoryear{Postill}{Postill}{2016}]%
        {postill2016doing}
\bibfield{author}{\bibinfo{person}{John Postill}.}
  \bibinfo{year}{2016}\natexlab{}.
\newblock \showarticletitle{Doing remote ethnography}.
\newblock In \bibinfo{booktitle}{\emph{The Routledge Companion to Digital
  Ethnography}}. \bibinfo{publisher}{Routledge}, \bibinfo{address}{London},
  \bibinfo{pages}{61--69}.
\newblock


\bibitem[\protect\citeauthoryear{R~Shamshiri, Weltzien, Hameed, J~Yule,
  E~Grift, Balasundram, Pitonakova, Ahmad, and Chowdhary}{R~Shamshiri
  et~al\mbox{.}}{2018}]%
        {shamshiri2018research}
\bibfield{author}{\bibinfo{person}{Redmond R~Shamshiri},
  \bibinfo{person}{Cornelia Weltzien}, \bibinfo{person}{Ibrahim~A Hameed},
  \bibinfo{person}{Ian J~Yule}, \bibinfo{person}{Tony E~Grift},
  \bibinfo{person}{Siva~K Balasundram}, \bibinfo{person}{Lenka Pitonakova},
  \bibinfo{person}{Desa Ahmad}, {and} \bibinfo{person}{Girish Chowdhary}.}
  \bibinfo{year}{2018}\natexlab{}.
\newblock \showarticletitle{Research and development in agricultural robotics:
  A perspective of digital farming}.
\newblock \bibinfo{journal}{\emph{International Journal of Agricultural and
  Biological Engineering}} \bibinfo{volume}{11}, \bibinfo{number}{4}
  (\bibinfo{date}{July} \bibinfo{year}{2018}), \bibinfo{pages}{1--14}.
\newblock


\bibitem[\protect\citeauthoryear{Rippey}{Rippey}{2019}]%
        {rippey_usda_2019}
\bibfield{author}{\bibinfo{person}{Brad Rippey}.}
  \bibinfo{year}{2019}\natexlab{}.
\newblock \bibinfo{title}{Nation's Wettest 12-Month Period on Record Slows Down
  2019 Planting Season}.
\newblock
\newblock
\urldef\tempurl%
\url{https://www.usda.gov/media/blog/2019/06/14/nations-wettest-12-month-period-record-slows-down-2019-planting-season}
\showURL{%
\tempurl}


\bibitem[\protect\citeauthoryear{US Citizenship and Immigration Services}{US
  Citizenship and Immigration Services}{2020}]%
        {USCIS2020}
US Citizenship and Immigration Services \bibinfo{year}{2020}\natexlab{}.
\newblock \bibinfo{title}{H-2A Temporary Agricultural Workers}.
\newblock
\newblock
\urldef\tempurl%
\url{https://www.uscis.gov/working-united-states/temporary-workers/h-2a-temporary-agricultural-workers}
\showURL{%
\tempurl}


\bibitem[\protect\citeauthoryear{Vogel, Kwiatkowski, and Violanti}{Vogel
  et~al\mbox{.}}{2011}]%
        {vogel_kwiatkowski_violanti_2011}
\bibfield{author}{\bibinfo{person}{Vogel}, \bibinfo{person}{Jane Kwiatkowski},
  {and} \bibinfo{person}{Anthony Violanti}.} \bibinfo{year}{2011}\natexlab{}.
\newblock \showarticletitle{On the job, in an empire built on eggs People Talk:
  A conversation with egg man Scott Kreher}.
\newblock \bibinfo{journal}{\emph{The Buffalo News}} (\bibinfo{date}{Apr}
  \bibinfo{year}{2011}).
\newblock
\urldef\tempurl%
\url{https://buffalonews.com/2011/04/24/on-the-job-in-an-empire-built-on-eggs-people-talk-a-conversation-with-egg-man-scott-kreher/}
\showURL{%
\tempurl}


\bibitem[\protect\citeauthoryear{Zuboff}{Zuboff}{2019}]%
        {zuboff2019age}
\bibfield{author}{\bibinfo{person}{Shoshana Zuboff}.}
  \bibinfo{year}{2019}\natexlab{}.
\newblock \bibinfo{booktitle}{\emph{The Age of Surveillance Capitalism: The
  Fight for a Human Future at the New Frontier of Power: Barack Obama's Books
  of 2019}}.
\newblock \bibinfo{publisher}{Profile Books}.
\newblock


\end{thebibliography}
\end{document}